\documentclass[prd,
,superscriptaddress
,nofootinbib,%
tightenlines ]{revtex4}
\usepackage{epsfig}
\usepackage{amsfonts}
\usepackage{color}
\usepackage{amssymb}

\newcommand{\ben}{\begin{displaymath}}
\newcommand{\een}{\end{displaymath}}
\newcommand{\be}{\begin{equation}}
\newcommand{\ee}{\end{equation}}
\newcommand{\bea}{\begin{eqnarray}}
\newcommand{\eea}{\end{eqnarray}}
\begin{document}
\title{Vacuum energy in the effective field theory of general relativity with a scalar field}

\author{E.~Epelbaum}
 \affiliation{Institut f\"ur Theoretische Physik II, Ruhr-Universit\"at Bochum,  D-44780 Bochum,
 Germany}
\author{J.~Gegelia}
 \affiliation{Institut f\"ur Theoretische Physik II, Ruhr-Universit\"at Bochum,  D-44780 Bochum,
 Germany}
 \affiliation{Tbilisi State
University, 0186 Tbilisi, Georgia}
\author{Ulf-G.~Mei{\ss}ner}
 \affiliation{Helmholtz Institut f\"ur Strahlen- und Kernphysik and Bethe
   Center for Theoretical Physics, Universit\"at Bonn, D-53115 Bonn, Germany}
 \affiliation{Institute for Advanced Simulation,  Forschungszentrum J\"ulich, D-52425 J\"ulich,
Germany}
\affiliation{Tbilisi State  University,  0186 Tbilisi, Georgia}
\author{L.~Neuhaus}
 \affiliation{Institut f\"ur Theoretische Physik II, Ruhr-Universit\"at Bochum,  D-44780 Bochum,
 Germany}

\begin{abstract}
A consistency condition of general relativity as an effective field
theory in Minkowskian background uniquely fixes the value of the
cosmological constant.  
In two-loop calculations, including the interaction of gravitons with matter fields, it has been shown that 
this value of the cosmological constant leads to vanishing vacuum energy, under the assumption that the
energy-momentum tensor of the gravitational field  is given by the pseudotensor of Landau-Lifshitz's classic textbook.
Here, we demonstrate that this result also holds when the self-interaction of a scalar field is taken into account. 
That is, our two-loop-order calculation suggests that in an effective
field theory of metric and scalar fields, one arrives at a consistent
theory with massless gravitons if the cosmological constant is fixed 
from the condition of vanishing vacuum energy. Vice versa, imposing
the consistency condition in Minkowskian background leads to a vanishing vacuum energy. 
\end{abstract}

\maketitle
	
It is widely accepted that at low energies, the physics of fundamental interactions is adequately described by effective field theory (EFT) \cite{Weinberg:1995mt,Weinberg:1996kr,Meissner:2022cbi}. 
Gravitation can also be included in this formalism by considering the most general effective Lagrangian of metric
and matter fields  \cite{Donoghue:1994dn,Donoghue:2015hwa,Donoghue:2017pgk}, which is invariant under all underlying
symmetries including the gauge symmetry of massless spin-two particles \cite{Veltman:1975vx}. 
It is well-known that for non-vanishing values of the cosmological
constant term, $\Lambda$, a quantum field theoretical treatment of
general relativity with the metric field presented as the Minkowskian 
background plus the graviton field poses a problem due to 
the graviton propagator possessing a pole that corresponds to  a massive ghost mode \cite{Veltman:1975vx}. 
Setting $\Lambda$ equal to zero does not improve the situation as
radiative corrections lead to the re-emergence of the problem with the massive ghost \cite{Burns:2014bva}. 
However, one can represent the cosmological constant as a
power series in $\hbar$ with coefficients chosen to yield
self-consistent EFT results to all orders in the loop expansion
\cite{Burns:2014bva}.   
Thus, the consistency requirement of a perturbative EFT in flat
Minkowski background uniquely fixes the cosmological constant 
as a function of other parameters of the effective Lagrangian. 
This also necessarily requires considering an EFT in a
curved background field if any other value of the
cosmological constant is assumed.  
In this case, the mass term of the graviton can be removed at classical level by
imposing the equations of motion with respect to the background 
graviton field  \cite{Gabadadze:2003jq}. To the best of our knowledge,
a systematic study beyond tree level has not been done due to the
lack of an EFT on non-Minkowskian background.  

In Refs.~\cite{Gegelia:2019fjx,Gegelia:2019zrv} it was found by
performing two-loop calculations for gravitational interactions only that the vacuum energy
vanishes exactly for the value of the cosmological constant
corresponding to that of Ref.~\cite{Burns:2014bva}, i.e.,  for the value 
that guarantees the vanishing of the graviton mass  and of the vacuum expectation value of the graviton field.
Provided that this result holds to all orders, also when including
interactions among matter fields, the uniquely fixed value of the
cosmological constant yielding a self-consistent perturbative EFT on Minkowkian
background could also be interpreted as a consequence of imposing the condition of vanishing vacuum energy. 

In this work we consider a simple EFT of general relativity with
metric and a scalar matter fields. We perform two-loop-order
calculations for the value of the cosmological constant leading to a  
consistent EFT in order to see if the vacuum energy vanishes when self-interactions of the scalar field are taken into account. 
Notice that while there does not exist a commonly accepted expression of the energy-momentum tensor for the
gravitational field (see, e.g., Refs.~\cite{Babak:1999dc,Butcher:2008rf,Szabados:2009eka,Butcher:2010ja,Butcher:2012th}), 
Refs.~\cite{Gegelia:2019fjx,Gegelia:2019zrv}  and also the current work use the definition of the energy-momentum pseudotensor (EMPT) and of the full four-momentum of the matter and
gravitational fields as given in the classic textbook \cite{Landau:1982dva}. 

\medskip
In the framework of the EFT of general relativity, the action is given via the most general effective
Lagrangian of gravitational and matter fields, which is invariant under 
general coordinate transformations and other underlying symmetries of the considered model,
\begin{equation}
S = \int d^4x \sqrt{-g}\, \left\{ {\cal L}_{\rm gr}(g)
+{\cal L}_{\rm m}(g,\psi)\right\} = \int d^4x \sqrt{-g}\, \left\{ \frac{2}{\kappa^2} (R-2\Lambda)+{\cal L}_{\rm gr,ho}(g)
+{\cal L}_{\rm m}(g,\psi)\right\} = S_{\rm gr}(g)+S_{\rm m}(g,\psi) \,.
\label{action}
\end{equation}
Here, $\kappa^2=32 \pi G$, with Newton's constant $G=6.70881\times 10^{-39}$ ${\rm
  GeV}^{-2}$, $\Lambda$ is the cosmological constant,
 $\psi$ and $g^{\mu\nu}$ denote the matter and metric fields, respectively,  $g=\det g^{\mu\nu}$, 
 and $R$ is the scalar curvature. An infinite number of gravitational self-interactions involving higher orders in derivatives  are contained in ${\cal L}_{\rm gr,ho}(g)$ 
 and ${\cal L}_{\rm m}(g,\psi)$ is the effective Lagrangian of the
 matter fields interacting with the metric and  the vielbein tetrad
 fields. Experimental evidence suggests that the contributions of
 ${\cal L}_{\rm gr,ho}(g)$ and of non-renormalizable interactions of
 ${\cal L}_{\rm m}(g,\psi)$  to physical quantities  are heavily
 suppressed. 

\medskip

The action of the matter part of the model considered here includes an
infinite number of terms, of which we show below only those
contributing in our specific  calculations: 
\begin{equation}
S_{\rm m} = \int d^4x \sqrt{-g}\, \left\{ 
\frac{g^{\mu\nu}}{2}\,\partial_\mu H \partial_\nu H -\frac{m^2}{2}\,H^2+\frac{g}{3!} \,H^3 + f H \right\},
\label{MAction}
\end{equation}
where 
$H$ is the scalar field. 
The low-energy  EFT is obtained by representing the metric field as the sum of the
Minkowskian background and the quantum fields \cite{tHooft:1974toh}
\begin{eqnarray}
g_{\mu\nu} &=& \eta_{\mu\nu}+\kappa h_{\mu\nu},\nonumber \\
g^ {\mu\nu} &=& \eta^{\mu\nu}-\kappa h^{\mu\nu}+\kappa^2 h^\mu_\lambda h^{\lambda\nu}-\kappa^3 h^\mu_\lambda h^{\lambda}_{\sigma} h^{\sigma\nu}+\cdots ~,
\label{gexpanded}
\end{eqnarray}
and we  calculate physical quantities perturbatively by expanding in $\kappa$ and other coupling constants treating them independently. 

\medskip

The energy-momentum tensor of the matter fields coupled to the gravitational field is given by
\begin{eqnarray}
T^{\mu\nu}_{\rm m} (g,\psi) & = & \frac{2}{\sqrt{-g}}\frac{\delta S_{\rm m} }{\delta g_{\mu\nu}}\,.
\label{EMTMatter}
\end{eqnarray}
The energy-momentum tensor corresponding to Eq.~(\ref{MAction}) has the form
\begin{eqnarray}
T_m^{\mu\nu}  =   
\partial_\mu H \partial_\nu H 
- g^{\mu\nu} \left\{ 
\frac{g^{\alpha\beta}}{2}\,\partial_\alpha H \partial_\beta H -\frac{m^2}{2}\,H^2 +\frac{g}{3!} \,H^3 + f H  \right\} \,.
\label{MEMT}
\end{eqnarray}
The pseudotensor
of the gravitational field (alone) is given as 
\begin{eqnarray}
T^{\mu\nu}_{\rm gr} (g)&=& 
\frac{4}{\kappa^2} \, \Lambda\,g^{\mu\nu} +  T_{LL}^{\mu\nu}(g)\,,
\label{defTs}
\end{eqnarray}
\noindent
where $T_{LL}^{\mu\nu}(g)$ is defined via \cite{Landau:1982dva}
\begin{eqnarray}
(-g)T^{\mu\nu}_{LL} (g) &=& \frac{2}{\kappa^2} \left(\frac{1}{8} \, g^{\lambda \sigma } g^{\mu \nu } g_{\alpha \gamma}
g_{\beta\delta} \, \mathfrak{g}^{\alpha \gamma},_{\sigma }
   \, \mathfrak{g}^{\beta \delta},_\lambda -\frac{1}{4} \, g^{\mu \lambda } g^{\nu\sigma } g_{\alpha ,\gamma} g_{\beta \delta }
  \, \mathfrak{g}^{\alpha\gamma},_\sigma \, \mathfrak{g}^{\beta\delta},_\lambda -\frac{1}{4} \, g^{\lambda \sigma } g^{\mu \nu } 
   g_{\beta \alpha} g_{\gamma \delta} \, \mathfrak{g}^{\alpha \gamma},_\sigma \, \mathfrak{g}^{\beta \delta},_\lambda \right.\nonumber\\
&+& \left. \frac{1}{2}\,  g^{\mu \lambda } g^{\nu\sigma } g_{\beta \alpha} g_{\gamma \delta } \, \mathfrak{g}^{\alpha \gamma},_\sigma \, \mathfrak{g}^{\beta\delta},_\lambda 
+g^{\beta \alpha } g_{\lambda \sigma }
  \, \mathfrak{g}^{\nu \sigma},_\alpha \, \mathfrak{g}^{ \mu\lambda},_\beta +\frac{1}{2} \, g^{\mu \nu } g_{\lambda \sigma }
   \, \mathfrak{g}^{\lambda \beta},_\alpha \, \mathfrak{g}^{\alpha\sigma},_\beta \right.\nonumber\\
&-& \left.g^{\mu \lambda } g_{\sigma \beta }
   \, \mathfrak{g}^{\nu \beta},_\alpha \, \mathfrak{g}^{\sigma\alpha},_\lambda 
   -g^{\nu \lambda } g_{\sigma \beta} \, \mathfrak{g}^{\mu\beta},_\alpha \, \mathfrak{g}^{\sigma \alpha},_\lambda
   +\, \mathfrak{g}^{\lambda \sigma},_\sigma \, \mathfrak{g}^{\mu\nu},_\lambda 
   - \, \mathfrak{g}^{\mu \lambda},_\lambda \, \mathfrak{g}^{\nu \sigma},_\sigma \right),
   \label{LLEMT}
\end{eqnarray}
with $\mathfrak{g}^{\mu\nu}=\sqrt{-g} \, g^{\mu\nu}$ and $\mathfrak{g}^{\mu\nu},_\lambda=\partial\mathfrak{g}^{\mu\nu}/\partial x^\lambda $.

The conserved full four-momentum of the matter and the gravitational fields is defined via the full EMPT $T^{\mu\nu}=T^{\mu\nu}_{\rm m} (g,\psi)+T^{\mu\nu}_{\rm gr} (g)$ as \cite{Landau:1982dva}
\begin{equation}
P^\mu= \int (-g) \, T^{\mu\nu} d S_\nu\,,
\label{EMV}
\end{equation}
where the integration covers any hypersurface containing the whole three-dimensional  space. Thus, for vanishing vacuum expectation value of $ (-g) \,  T^{\mu\nu} $,
the energy of the vacuum will be zero.
This expectation value can be represented via the following path integral:
\begin{eqnarray}
\langle 0| (-g) T^{\mu\nu} |0\rangle &=& 
\int {\cal D }g\, {\cal D}\psi  \, (-g)\left[ T^{\mu\nu}_{\rm gr}(g)+ T_{\rm m}^{\mu\nu}(g,\psi) \right] \exp\left\{ i \int d^4 x \, \sqrt{-g}\,\left[ {\cal L}(g,\psi) +{\cal L}_{\rm GF}\right] \right\},
\label{VacuumE}
\end{eqnarray}
where 
\begin{equation}
{\cal L}_{\rm GF}= \xi \left( \partial_\nu h^{\mu\nu}-\frac{1}{2} \partial^\mu h^\nu_\nu \right) \left( \partial^\beta h_{\mu\beta}-\frac{1}{2} \partial_\mu h^\alpha_\alpha \right)
\label{GFT}
\end{equation}
 is the gauge-fixing term with parameter $\xi$, and  the integration measure includes the Faddeev-Popov determinant.

\medskip   

\begin{figure}[t]
  \begin{center}
   \includegraphics[height=7.0cm]{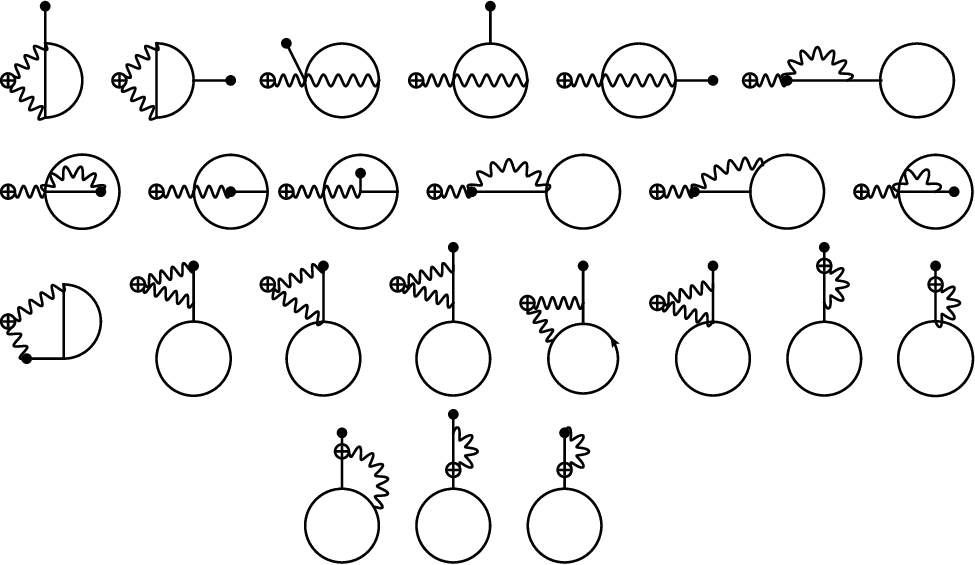}  
    \caption{Topologies of two-loop diagrams whose sums  (but not the separate diagrams) lead to the same contribution to the cosmological constant for the graviton tadpole and the matrix element of  $ (-g) \,  T^{\mu\nu} $. 
    The wavy and solid lines correspond to gravitons and scalars, respectively. The filled dots stand for vertices generated by the $f H$ term of the Lagrangian. 
    The cross corresponds to either external graviton line or a $ (-g)
    \,  T^{\mu\nu} $ insertion.
}
\label{EMTLoop}
\end{center}
\end{figure}

According to Ref.~\cite{Burns:2014bva} the cosmological constant
$\Lambda$ is uniquely fixed by the condition that the vacuum
expectation value of the quantum field $h_{\mu\nu}$ vanishes, and this also
 removes the graviton mass from the dressed propagator as a result of a Ward identity \cite{Burns:2014bva}. 
To study the implications for the vacuum energy at two loop order we represent $\Lambda$  as 
\begin{equation}
\Lambda = \sum_{i=0}^\infty \hbar^i \Lambda _i \,.
\label{CCexpanded}
\end{equation}
The vacuum expectation value of
$(-g)T^{\mu\nu}$
at tree order vanishes for $\Lambda_0=0$, and
this choice also removes the graviton mass from the propagator at tree level. 
It is a straightforward consequence of Eq.~(\ref{EMTMatter}) that the same value of $\Lambda_1$ cancels the one-loop contribution
to the vacuum expectation value of the graviton field $h_{\mu\nu}$ and the vacuum energy.
The first non-trivial result is obtained at two-loop order. Calculations taking into account only gravitational interaction have been done in Refs.~\cite{Gegelia:2019fjx,Gegelia:2019zrv}.
In this work we consider the order-$fg$ two-loop contributions to the
vacuum expectation values of $ (-g) \,  T^{\mu\nu} $ and of the graviton field $h_{\mu\nu}$,  and we
find that the same value of $\Lambda_2$ leads to an exact cancelation of both of them.\footnote{In
  calculations we used the program FeynCalc
  \cite{Mertig:1990an,Shtabovenko:2016sxi}.}  
Again, it is a trivial consequence of Eq.~(\ref{EMTMatter}) that
separate diagrams with the same structure, where the vertices with
external graviton and with $ (-g) \,  T^{\mu\nu} $ insertion couple
only to scalar lines, give equal $\Lambda_2$-contributions. On the
other hand, only the sums of diagrams of the topologies shown in
Fig.~\ref{EMTLoop} lead to
identical expressions when imposing the two conditions. 
That is, we first fix $\Lambda$ by imposing on the corresponding
diagrams the condition that the expectation value of $h_{\mu\nu}$
vanishes, and next by imposing on corresponding diagrams the condition
that the vacuum expectation value of $ (-g) \,  T^{\mu\nu} $
vanishes. We find that individual diagrams contributing to a given topology in
Fig.~\ref{EMTLoop} yield different contributions to the cosmological
constant $\Lambda_2$ in the two cases, but considering the sums of the two sets of
diagrams results in the same value of the cosmological constant.

\medskip

\medskip

A self-consistent EFT should lead to finite physical quantities after renormalizing (an infinite number of)
parameters of the effective Lagrangian. 
Therefore, it is mandatory that the unique value of the cosmological
constant that defines the
perturbative EFT of the Standard Model, coupled to gravitons on a Minkowskian flat background, leads to a finite expression of the 
energy (density) of the vacuum to all orders of perturbation theory. 
Based on the two-loop order results including the one of the current
work, we expect that this finite value should be at least of three-loop-order, and thus very small. 
Moreover, it seems natural that an adequate theory formulated in the language of mathematics should assign zero energy to the vacuum state \cite{Faddeev:1982id}.
Therefore turning the argument around we expect that by demanding that
the vacuum energy should be vanishing to all orders  
we obtain a self-consistent perturbative low-energy EFT of matter and gravitational fields on a flat Minkowskian background.

\acknowledgments

We thank D.~Djukanovic for checking the completeness of the list of
diagrams and their symmetry factors and for useful discussions.  
This work was supported in part by BMBF (Grant No. 05P21PCFP1), by
DFG and NSFC through funds provided to the Sino-German CRC 110
“Symmetries and the Emergence of Structure in QCD” (NSFC Grant
No. 11621131001, DFG Project-ID 196253076 - TRR 110), by the Georgian Shota Rustaveli National
Science Foundation (Grant No. FR-23-856),
by CAS through a President's International Fellowship Initiative (PIFI)
(Grant No. 2018DM0034), by the EU Horizon 2020 research and
innovation program (STRONG-2020, grant agreement No. 824093),
and by the Heisenberg-Landau Program 2021.


\begin{references}

\bibitem{Weinberg:1995mt}
S.~Weinberg,
``The Quantum theory of fields. Vol. 1: Foundations,''
Cambridge University Press, 1995.

\bibitem{Weinberg:1996kr}
S.~Weinberg,
``The quantum theory of fields. Vol. 2: Modern applications,''
Cambridge University Press, 1996.
  
 \bibitem{Meissner:2022cbi}
U.-G.~Mei{\ss}ner and A.~Rusetsky,
``Effective Field Theories,''
Cambridge University Press, 2022.

\bibitem{Donoghue:1994dn} 
  J.~F.~Donoghue,
  Phys.\ Rev.\ D {\bf 50}, 3874 (1994).

  

\bibitem{Donoghue:2015hwa} 
  J.~F.~Donoghue and B.~R.~Holstein,
  J.\ Phys.\ G {\bf 42}, no. 10, 103102 (2015).
 
\bibitem{Donoghue:2017pgk} 
  J.~F.~Donoghue, M.~M.~Ivanov and A.~Shkerin,
  arXiv:1702.00319 [hep-th].
 
\bibitem{Veltman:1975vx} 
  M.~J.~G.~Veltman,
  Conf.\ Proc.\ C {\bf 7507281}, 265 (1975).

\bibitem{Burns:2014bva} 
  D.~Burns and A.~Pilaftsis,
  Phys.\ Rev.\ D {\bf 91}, no. 6, 064047 (2015).


%
%
%

\bibitem{Gabadadze:2003jq} 
  G.~Gabadadze and A.~Gruzinov,
  Phys.\ Rev.\ D {\bf 72}, 124007 (2005).

\bibitem{Gegelia:2019fjx}
J.~Gegelia and U.-G.~Mei\ss{}ner,
Phys. Rev. D \textbf{100}, 
046021 (2019).

\bibitem{Gegelia:2019zrv}
J.~Gegelia and U.-G.~Mei\ss{}ner,
Phys. Rev. D \textbf{100}, 
124002 (2019).



\bibitem{Babak:1999dc} 
  S.~V.~Babak and L.~P.~Grishchuk,
  Phys.\ Rev.\ D {\bf 61}, 024038 (2000).

\bibitem{Butcher:2008rf} 
  L.~M.~Butcher, A.~Lasenby and M.~Hobson,
  Phys.\ Rev.\ D {\bf 78}, 064034 (2008):

\bibitem{Szabados:2009eka} 
  L.~B.~Szabados,
  Living Rev.\ Rel.\  {\bf 12}, 4 (2009).

\bibitem{Butcher:2010ja} 
  L.~M.~Butcher, M.~Hobson and A.~Lasenby,
  Phys.\ Rev.\ D {\bf 82}, 104040 (2010).
  
\bibitem{Butcher:2012th} 
  L.~M.~Butcher, M.~Hobson and A.~Lasenby,
  Phys.\ Rev.\ D {\bf 86}, 084013 (2012).
  
\bibitem{Landau:1982dva} 
  L.~D.~Landau and E.~M.~Lifschits,
  ``The Classical Theory of Fields,'' Oxford: Pergamon Press (1975).

\bibitem{tHooft:1974toh} 
  G.~'t Hooft and M.~J.~G.~Veltman,
  Ann.\ Inst.\ H.\ Poincare Phys.\ Theor.\ A {\bf 20}, 69 (1974).

  
\bibitem{Mertig:1990an}
  R.~Mertig, M.~Bohm and A.~Denner,
  Comput.\ Phys.\ Commun.\  {\bf 64}, 345 (1991).

\bibitem{Shtabovenko:2016sxi}
  V.~Shtabovenko, R.~Mertig and F.~Orellana,
  Comput.\ Phys.\ Commun.\  {\bf 207} (2016) 432

\bibitem{Faddeev:1982id}
L.~D.~Faddeev,
Sov. Phys. Usp. \textbf{25}, 130-142 (1982).

\end{references}
\end{document}